\documentclass[english,12pt,a4]{article}
\usepackage[T1]{fontenc}
\usepackage{babel}
\usepackage{refstyle}
\usepackage{textcomp}
\usepackage[font={footnotesize},margin=1.5cm]{caption}
\usepackage{amsmath}
\usepackage{amssymb}
\usepackage{graphicx}
\usepackage[a4paper, top=4.5cm,bottom=2.5cm,left=2.5cm,right=2cm]{geometry}
\usepackage{float}
\begin{document}

\title{A new analysis of quasar polarisation alignments}
\author{V.~Pelgrims\thanks{pelgrims@astro.ulg.ac.be}\ 
~and J.R. Cudell\thanks{jr.cudell@ulg.ac.be}\\
{\small IFPA, AGO dept., University of Li\`ege, B4000 Li\`ege, Belgium}\\}
\maketitle
\begin{abstract}
We propose a new method to analyse the alignment of optical polarisation
vectors from quasars. This method leads to a definition of intrinsic preferred axes and to a determination of the probability $p^{\sigma}$ that the distribution of polarisation directions is random. This probability is found to be as low as $3.0\, 10^{-5}$ for one of the regions of redshift.
\end{abstract}

\section{Introduction}
\label{sec:Intro}
Fifteen years ago, Hutsem\'ekers discovered that the optical polarisation
vectors from distant quasars are not randomly distributed \cite{key-1}.
The original study considered a sample of 170 quasars, and the probability
of a random distribution was 0.5 \%. Since then, further measurements
have been added \cite{key-2,key-10,key-3}, and the significance
of the effect has grown: with the present sample of 355 polarisation
data points, the alignment effect has been estimated to have a probability
to be random lower than 0.1 \% in some regions of the sky defined by cuts on right
ascension and redshift.

The original test however had a weak point, as the angles of the polarisation
vectors are measured with respect to their local meridian. Therefore,
the strength of the alignment, and the significance of the effect,
depend on the choice of the spherical coordinate system, so that it
is possible to find an axis for which the effect becomes negligible,
i.e. for which the data look random. Indeed, the original statistical
tests do not take into account the fact that polarisation vectors
come from different lines of sight and thus are not defined in the
same plane. To remedy this problem, \cite{key-6} have proposed
to parallel-transport the polarisation vectors along geodesics of
the celestial sphere onto one point in order to compare them. While
the distribution of polarisation angles still depends on the point
where it is built, it does not depend on the coordinates, and one
can build robust statistical tests (see Jain et al. (2004) and Hutsem\'ekers et al. (2005)).
Unfortunately, these tests have to resort to the generation of a large
number of random data in order to determine the significance of the
signal and do not lead to a clear characterization of the effect.
Furthermore, to make an intrinsic test, polarisation vectors must
be transported to the location of the quasars. However, this is an
arbitrary choice as they can be transported to any point on the celestial
sphere. The sum of the parallel-transported polarisation vectors then
makes a continuous axial vector field on the sphere, and by the hairy
ball theorem \cite{hbt}, it is still possible to find at least two
points where the effect vanishes. Hence the quantification of the
size of the effect seems uncertain again.

We propose here another method which is totally independent of the
coordinate system, and which quantifies unambiguously the alignment
effect. It allows us to compare the polarisation vectors of sources
located at different angular coordinates and leads to the characterization
of the effect through a blind analysis of the data. The basic idea
is to consider the physical polarisation vectors as 3-dimensional
objects rather than 2-dimensional ones embedded in their polarisation
plane. These 3-dimensional objects are the directions of the electric
field oscillations and they are the physical objects which are measured.
As we are dealing with a number of vectors, it is clear that it will
be possible to define preferred directions. We can also use the same
method to study the dependence on redshift, position in the sky or degree of linear polarisation by imposing cuts on these variables and repeating
the study for the corresponding sub-sample.

We devote the present paper to the construction of this new statistical
method applied here to the analysis of quasar polarisation data, in
which the evaluation of the significance of the signal will be largely
analytic. The second section of this paper explains the details of
our statistical method. It also contains illustrations and discussions
related to the statistical background, when polarisation angles are
assumed to be uniformly distributed, and is compared with the original
study of \cite{key-3} using the same cuts.
The third section is devoted to the application of this new statistical
method first globally, then to slices in redshift. We also consider
there the dependence of the alignment on the various parameters, the
possibility of a cosmological alignment, and define our final most
significant regions exhibiting an anomalous alignment of polarisation
vectors.

\section{A coordinate-invariant statistical test for polarisation data}
\label{sec:coordInvTest}
When an electromagnetic wave is partially or fully linearly polarized,
a polarisation vector is introduced. Its norm reflects the degree
of linear polarisation of the radiation while its direction is that
of the oscillating electric field. This vector is thus embedded into
the plane orthogonal to the radiation direction of propagation, the
polarisation plane. Since the electric field is oscillating, the polarisation
vector is an axial quantity, rather than a true vector, so that the
polarisation angle is determined up to $\pi$ radians.

We consider sources as being points on the unit celestial sphere and
we choose a spherical polar coordinate system defined by the orthonormal
3-vectors $(\boldsymbol{e}_{r},\,\boldsymbol{e}_{\theta},\,\boldsymbol{e}_{\phi})$,
with $\boldsymbol{e}_{\theta}$ pointing to the South pole. In the
following bold-faced letters indicate 3-vectors. Polarisation vectors
are tangent to this unit sphere. For a given source in the direction
$\boldsymbol{e}_{r}$, a polarisation vector must lie in the plane
defined by the two unit vectors $\boldsymbol{e}_{\phi}$ and $\boldsymbol{e}_{\theta}$.
We choose the angle $\psi$ between the polarisation vector $\boldsymbol{p}$
and the basis vector $\boldsymbol{e}_{\phi}$, defined in the range
$[-\pi/2,\,\pi/2[$, to be the polarisation angle. The normalized
polarisation vector can then be written
\begin{equation}
\hat{\boldsymbol{p}}=\mbox{cos}\psi\,\boldsymbol{e}_{\phi}-\mbox{sin}\psi\,\boldsymbol{e}_{\theta}\;.\label{eq:1}
\end{equation}
Each measurement $(i)$ of the dataset \cite{key-3}
is equivalent to a position 3-vector $\boldsymbol{e}_{r}^{(i)}$associated
with a normalised polarisation direction $\hat{\boldsymbol{p}}^{(i)}$
and polarisation magnitude $|\boldsymbol{p}^{(i)}|$. Contrarily to
the various angles, $\boldsymbol{e}_{r}^{(i)}$ and $\hat{\boldsymbol{p}}^{(i)}$
are physical, $i.e.$ they do not depend on the choice of coordinates.
As we are interested in polarisation alignments, we shall consider
mostly the $\hat{\boldsymbol{p}}^{(i)}$ in the following. 

\begin{figure}
\begin{center}
%\begin{minipage}{126mm}
\includegraphics[scale=0.4]{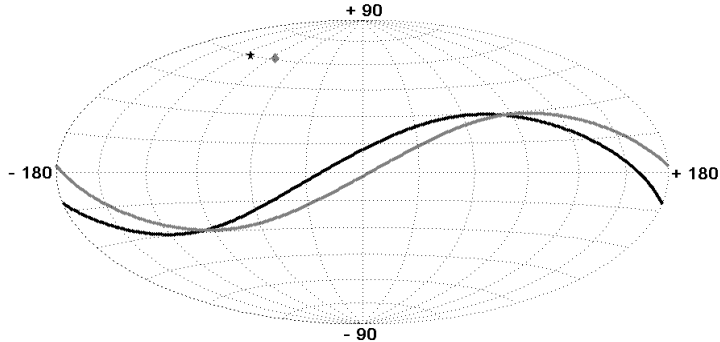}
\caption{Superposition of the Hammer-Aitoff
projections of the celestial sphere and the polarisation sphere (in
galactic coordinates). Two quasars (B1115+080 (in black) and B1157+014
(in grey)) are displayed on the celestial sphere with the corresponding
geometric loci of their polarisation point on the polarisation sphere.
The source position and the corresponding geometric locus of the polarisation
point are printed in same brightness.}
\label{fig:2Qexample}
%\end{minipage}
\end{center}
\end{figure}

The problem is then as follows: we have a number of normalised vectors,
and we want to decide if they are abnormally aligned. We can draw
them from the same origin, and their ends, which we shall call the
polarisation points, have to lie on a unit 2-sphere, which we shall
refer to as the polarisation sphere. The problem is that, even when the polarisation angles $\psi$ are uniformly distributed, the polarisation sphere is not uniformly covered by the points: they have to lie on great circles on the 2-sphere. Indeed, for each source, the
polarisation vectors are constrained to be in the plane defined by
the basis vectors $(\boldsymbol{e}_{\phi}^{(i)},\,\boldsymbol{e}_{\theta}^{(i)})$.
The intersection of the plane with the polarisation sphere is
a great circle, which is the geometric locus where the polarisation
vector attached to the source $(i)$ may intersect the sphere, as
shown in Fig. \ref{fig:2Qexample}. 
Note that the figure
is symmetric as polarisation vectors are defined up to a sign. In
the following, we choose to show the full sphere, although a half-sphere
could be used to represent the polarisation space.

To compare the observations with what one would expect if the polarisation points were drawn from a random distribution of polarisation angles, we need to select a region on the polarisation sphere, count the number of polarisation points within this region, and compare this number with the prediction.
One could do this by Monte-Carlo techniques, but as we shall see the probabilities turn out to be rather low, so that a detailed study would prove difficult.

However, we found that a particular choice of shape for the region
on the sphere considerably simplifies the evaluation of probabilities.
We consider cones in which the polarisation vectors fall, or equivalently
spherical caps of fixed aperture angle. The probability distribution
of a given number of points in a given spherical cap can be computed
analytically as explained below. A scan of the whole polarisation
half-sphere leads to a map of expected densities which constitutes
the statistical background. At any location on the half-sphere, the
hypothesis of uniformity can then be tested by calculating the probability
of the observed number of polarisation points. An alignment of polarisation
vectors from different sources will be detected when an over-density
between data points and the background is significant.

\subsection{Construction of the probability distribution}
\label{subsec:PnDistConst}
As mentioned above, the locus of the polarisation points is a half-circle
in the plane normal to the source position vector. The probability
that a polarisation point lies inside a spherical cap is then given
by the length of the arc of circle intercepted by the cap, divided
by the whole length of the half-circle ($\pi$). Let $\eta$ being
the half-aperture angle of the spherical cap, and $\hat{\boldsymbol{s}}$
the unit vector pointing to its centre. If $\hat{\boldsymbol{p}}^{(i)}$
is a normalised polarisation vector attached to the source $(i)$,
with position vector $\boldsymbol{e}_{r}^{(i)}$, the corresponding
polarisation point lies inside the spherical cap centred at $\hat{\boldsymbol{s}}$
if and only if

\begin{equation}
|\hat{\boldsymbol{p}}^{(i)}\cdot\hat{\boldsymbol{s}}|\geq\mbox{cos}\eta\label{eq:1b}
\end{equation}
is verified. Adopting the decomposition of $\hat{\boldsymbol{p}}^{(i)}$ along two vectors in the polarisation plane
\begin{equation}
\hat{\boldsymbol{p}}^{(i)}=A\, \left(\hat{\boldsymbol{s}}-\left(\hat{\boldsymbol{s}}\cdot\boldsymbol{e}_{r}^{(i)}\right)\boldsymbol{e}_{r}^{(i)} \right) + B\, \hat{\boldsymbol{t}}^{(i)}  
\label{eq:2}
\end{equation}
where $\hat{\boldsymbol{t}}^{(i)}=\left(\boldsymbol{e}{}_{r}^{(i)}\times\hat{\boldsymbol{s}}\right)/|\boldsymbol{e}_{r}^{(i)}\times\hat{\boldsymbol{s}}|$,
a straightforward calculation involving the normalisation of $\hat{\boldsymbol{p}}^{(i)}$ and the condition for being
inside the spherical cap leads to the arc length $L^{(i)}$ of the
geometric locus lying inside the considered area. The result takes
a simple form in terms of $\tau^{(i)}\in[0,\,\pi[$, the angle between
$\boldsymbol{e}_{r}^{(i)}$ and $\hat{\boldsymbol{s}}$: condition (\ref{eq:1b})
becomes $\mbox{sin}\tau^{(i)}\geq\mbox{cos}\eta$ and, by integration of the line element, the arc length within the cap is found to be:
\begin{equation}
L^{(i)}=\begin{cases}
2\,\mbox{acos}\left(\frac{\cos\eta}{\sin\tau^{(i)}}\right) & \mbox{if}\:\sin\tau^{(i)}\geq\cos\eta\\
0 & \mbox{otherwise}
\end{cases}\;.\label{eq:3}
\end{equation}
\newcommand{\ppi}{\ell}
Therefore, the probability $\ppi^{(i)}$ that the \textit{i}-th source
of the sample leads to a polarisation point inside a given spherical cap
is: 
\begin{equation}
\ppi^{(i)}=\frac{L^{(i)}}{\pi}.\label{eq:4}
\end{equation}
Hence, these probabilities only depend on the
chosen aperture angle of the spherical cap and on the angle between
the source positions and the cap centre. These probabilities are thus
completely independent of the system of coordinates.

For each cap, the set of probabilities $\ppi^{(i)}$ leads to the
construction of the probability distribution $P_{n}$ of observing
exactly $n$ points of polarisation inside the spherical cap. If $N$
is the sample size we have: 
\begin{eqnarray}
P_{0} & = & \prod_{i=1}^{N}\left(1-\ppi^{(i)}\right)\label{eq:5}\\
P_{1} & = & \sum_{j=1}^{N}\ppi^{(j)}\,\prod_{i\neq j}\left(1-\ppi^{(i)}\right)\label{eq:6}\\
P_{2} & = & =\frac{1}{2}\,\sum_{k=1}^{N}\ppi^{(k)}\,\sum_{j\neq k}\ppi^{(j)}\,\prod_{i\neq j\neq k}\left(1-\ppi^{(i)}\right)\label{eq:7}\\
 & \vdots\nonumber \\
P_{N} & = & \frac{1}{N!}\,\sum_{l=1}^{N}\ppi^{(l)}\,\ldots\,\sum_{j\neq prev.indices}\ppi^{(j)}\,\prod_{i\subset\{\emptyset\}}\left(1-\ppi^{(i)}\right)\nonumber \\
 	 & = & \prod_{l=1}^{N}\ppi^{(l)}\quad.\label{eq:8}
\end{eqnarray}
Note that following the previous definitions, it is possible to write,
for each $n$,
\begin{equation}
P_{n}=\frac{1}{n}\,\sum_{j=1}^{N}\ppi^{(j)}\, P_{n-1}{}_{\setminus j}\label{eq:9}
\end{equation}
where $P_{n-1}{}_{\backslash j}$ is the probability to observe $n-1$
points of polarisation (and only $n-1$) after the \textit{j}-th element
is removed from the original sample, making the new sample size $N-1$.

\subsection{A fast algorithm for generating the $P_{n}$}
\label{PnDistAlgo}
\newcommand{\pp}{\rlap{\large /}{\ell}}

Starting with the entire sample of size $N$, let us consider the probability
$P_{0}$ to observe no polarisation point within the cap. We remove
the \textit{k}-th element from this sample. Then, from \eqref{5},
the probability to observe no polarisation point within this reduced
sample, denoted by $P_{0}{}_{\setminus k}$, is related to $P_{0}$
through $P_{0}=\pp^{(k)}\, P_{0}{}_{\setminus k}$, where we introduced
the following notation for the probability that the source \textit{$k$}
does not lead to a polarisation point in the concerned area : ${\pp^{(k)}}\equiv\left(1-\ppi^{(k)}\right)$. 

First consider the probability $P_{1}$ to observe one and only one
polarisation point:
\begin{eqnarray}
P_{1} & = & \sum_{j=1}^{N}\ppi^{(j)}\,\prod_{i\neq j}{\pp^{(i)}}\nonumber \\
 & = & \sum_{j=1}^{N}\ppi^{(j)}\, P_{0}{}_{\backslash j}\nonumber \\
 & = & \sum_{j\neq k}\frac{\ppi^{(j)}}{{\pp^{(j)}}}\, P_{0}+\ppi^{(k)}\, P_{0}{}_{\backslash k}\nonumber \\
 & = & \pp^{(k)}\,\left(\sum_{j\neq k}\frac{\ppi^{(j)}}{\pp^{(j)}}\, P_{0}{}_{\backslash k}\right)+\ppi^{(k)}\, P_{0}{}_{\backslash k}\nonumber \\
 & = & \pp^{(k)}\, P_{1}{}_{\backslash k}+\ppi^{(k)}\, P_{0}{}_{\backslash k}\quad.\label{eq:10}
\end{eqnarray}
A similar calculation leads to $P_{2}=\pp^{(k)}\, P_{2}{}_{\backslash k}+\ppi^{(k)}\, P_{1}{}_{\backslash k}$.
One can prove by induction that the following relation holds:
\begin{equation}
P_{m}=\pp^{(k)}\, P_{m}{}_{\backslash k}+\ppi^{(k)}\, P_{m-1}{}_{\backslash k}
\label{eq:11}
\end{equation}
Indeed, assuming the relation is true for $m\leq n-1$, it is easy to show
that it is then true for $m=n$:
\begin{eqnarray}
P_{n} & = & \frac{1}{n}\,\sum_{l=1}^{N}\ppi^{(l)}\, P_{n-1}{}_{\backslash l}\nonumber\\
 & = & \frac{1}{n}\,\sum_{l\neq k}\ppi^{(l)}\, P_{n-1}{}_{\backslash l}+\frac{1}{n}\,\ppi^{(k)}\, P_{n-1}{}_{\backslash k}\nonumber\\
  & = & \pp^{(k)}\,\left(\frac{1}{n}\,\sum_{l\neq k}\ppi^{(l)}\, P_{n-1}{}_{\backslash l\backslash k}\right)\nonumber\\
 & & +\ppi^{(k)}\,\left(\frac{1}{n}\,\sum_{l\neq k}\ppi^{(l)}\, P_{n-2}{}_{\backslash l\backslash k}\right)+\frac{1}{n}\,\ppi^{(k)}\, P_{n-1}{}_{\backslash k}\nonumber\\
 & = & \pp^{(k)}\, P_{n}{}_{\backslash k}\nonumber\\
 & & + \ppi{(k)}\, \left(\frac{n-1}{n}\, \left[\frac{1}{n-1}\,\sum_{l \neq k}\ppi^{(l)}\, P_{n-2}{}_{\backslash l\backslash k} \right] \right) \nonumber\\
 & & + \ppi{(k)}\,\frac{1}{n}\,P_{n-1}{}_{\backslash k}\nonumber\\
 & = & \pp^{(k)}\, P_{n}{}_{\backslash k}+\ppi^{(k)}\, P_{n-1}{}_{\backslash k}\quad.
\end{eqnarray}
Therefore, \eqref{11} holds%
\footnote{Equation (\ref{eq:11}) was first introduced by Howard (1972) and
its numerical behaviour was extensively discussed in the paper
\cite{key-9} about computational techniques for
the Poisson-binomial probabilities. The algorithm presented here is equivalent to that
given in \cite{key-9}.} for every $m\in\mathbb{N}$.
Assuming that, for a sample of size $N$ and for a given spherical
cap, we have all elementary probabilities $\ppi^{(i)}$, we use
the following algorithm for numerically computing the probability distribution:

1) We introduce a column vector $V$ of size $N+1$, initialized to
zero, except for $V_{0}$ which is set to 1. The $V_{n}$ are the
$P_{n}$ for an empty data set.

2) We add one data point at a time, and update $V$ according to \eqref{11}.

3) After $N$ iterations, the $V_{n}$ give the $P_{n}$ distribution
for the studied sample%
\footnote{We have tested our implementation of the algorithm by comparing its
results to those obtained via a Monte-Carlo treatment, for the A1
region of reference \cite{key-3}. We produced 10000 random samples,
for which the quasar positions were kept fixed and for which the polarisation
angles were randomly generated according to a uniform distribution.
For different spherical caps we built the corresponding $P_{n}$ distributions,
and found a perfect agreement. We checked that the same conclusions
are obtained for the whole sample of quasars presented in \cite{key-3}
and for arbitrary sub-samples. %
}.

\subsection{A first example}
\label{subsec:StatBackground}
\begin{figure}
\begin{center}
%\begin{minipage}{126mm}
%\begin{center}
\includegraphics[scale=0.4]{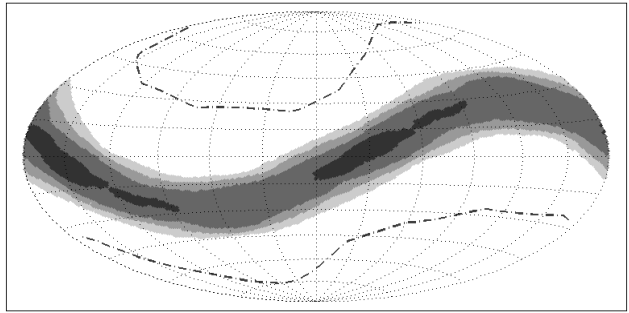}
%\end{center}
\caption{Hammer-Aitoff projection (galactic
coordinates) of the polarisation sphere associated to the A1 region.
Expected density regions are displayed following the legend: white:
$\bar{N^{a}}<4$; light grey: $4\leq\bar{N^{a}}<5$; grey: $5\leq\bar{N^{a}}<6$;
dark grey: $6\leq\bar{N^{a}}<7$; black: $\bar{N^{a}}\geq7$. White
regions towards poles which are delimited by dashed curves are regions
where polarisation points cannot fall at all.}
\label{fig:DensityRegionA1}
%\end{minipage}
\end{center}
\end{figure}

To illustrate the use of the above, we show in Fig. \ref{fig:DensityRegionA1} a map of the expected background for region A1 of \cite{key-3}, as
defined in Table \ref{tab:A123Huts}.

\begin{table}[H]
\begin{center}
\begin{tabular}{lcccc}
\hline 
region & declination & right ascension & redshift & number of quasars\tabularnewline
\hline
A1 & $\delta\leq50\text{\textdegree}$ & $168\text{\textdegree}\leq\alpha\leq217\text{\textdegree}$, & $1.0\leq z\leq2.3$ & 56\tabularnewline
A2 & $\delta\leq50\text{\textdegree}$ & $150\text{\textdegree}\leq\alpha\leq250\text{\textdegree}$ & $0.0\leq z<0.5$ & 53\tabularnewline
A3 &  & $320\text{\textdegree}\leq\alpha\leq360\text{\textdegree}$ & $0.7\leq z\leq1.5$ & 29\tabularnewline
\hline 
\end{tabular}
\caption{The three regions of alignment of in equatorial coordinates B1950.}
\label{tab:A123Huts}
\end{center}
\end{table}
At each point $a$ of the polarisation sphere we associate a probability
distribution $P_{n}^{a}$ through the use of spherical caps.
The mean values $\bar{N^{a}}=\sum_{n}nP_{n}^{a}$ determine the most
expected number of polarisation points. From those numbers, we build
iso-density regions on the polarisation sphere in order to visualize
the structure that the statistical background takes. We arbitrarily
choose here caps of half aperture $\eta=17\text{\textdegree}$. The
dependence of the results on $\eta$ will be discussed in section \ref{subsubsec:depEta}.

Due to the non uniformity of the source locations, there are regions
of maxima (and minima) in the expected densities of polarisation points
as well as regions where polarisation points are forbidden. For this
sample, a close look at Fig. \ref{fig:DensityRegionA1} shows
that a quadrupole is naturally expected in the density structure on
the polarisation sphere. This shows that the use of the $P_{n}$ distributions
is mandatory, as the expected density is not flat.

\subsection{Further refinements of the method}

\subsubsection{Optimal set of centres for the spherical caps}
\label{subsubsec:Border effect}

The method presented so far has two problems:
\begin{itemize}
\item Several spherical caps can contain the same polarisation points, so
that several probability distributions are assigned to the same set
of data points. 
\item Among the caps containing the same data points, the most significant
ones will be those for which several of the $\ppi^{(i)}$ will be
small, i.e. for which the loci of several polarisation points are
almost tangent to the caps. This enhanced significance is an artefact
of our method.
\end{itemize}
In order to minimize these problems, we do not allow all caps to be
considered, but rather focus on those that correspond to cones with
an axis along the vectorial sum of the normalised polarisation vectors
inside them. Hence the effective polarisation vector corresponding to the centre of the cap is 
\[
\boldsymbol{s}_{centre}=\sum_{i\in cap}\hat{\boldsymbol{p}}^{(i)},\:\hat{\boldsymbol{s}}_{centre}=\frac{\boldsymbol{s}_{centre}}{|\boldsymbol{s}_{centre}|}
\]
These centres are first determined by iteration before applying the
algorithm explained above.

\subsubsection{Local p-value of the data}
\label{subsubsec:pvalue}
The study of alignments is performed separately for each cap $a$
on the polarisation sphere, for which we derive probability distributions
$P_{n}^{a}$. In each cap, we count the number $o_{a}$ of observed
polarisation points, and $P_{o_{a}}^{a}$ gives us the probability
that the presence of $o_{a}$ polarisation points in cap $a$ is due
to a background fluctuation. The probability that a generation from
a uniform background has a density greater than the observed one is
given by the p-value $p^{a}=\,\sum_{n\geq o_{a}}\, P_{n}^{a}$. The
latter quantity gives us the significance level of a specific polarisation
point concentration in one given direction. As already mentioned,
\eqref{4} shows that $p$ is coordinate invariant. It in
fact provides a generalisation of the binomial test used in \cite{key-1,key-2,key-3}.
For each sample, we can consider the cap $a_{min}$ that gives the
most significant p-value $p_{min}=min_{a}(p^{a})$ which we shall
call the significance level. This defines a direction in polarisation
space, and a plane in position space.

\subsubsection{Dependence on the spherical cap aperture}
\label{subsubsec:depEta}
The only free quantity in this method is the aperture half-angle
of the spherical caps.

\begin{figure}[H]
\begin{center}
\includegraphics[scale=0.45]{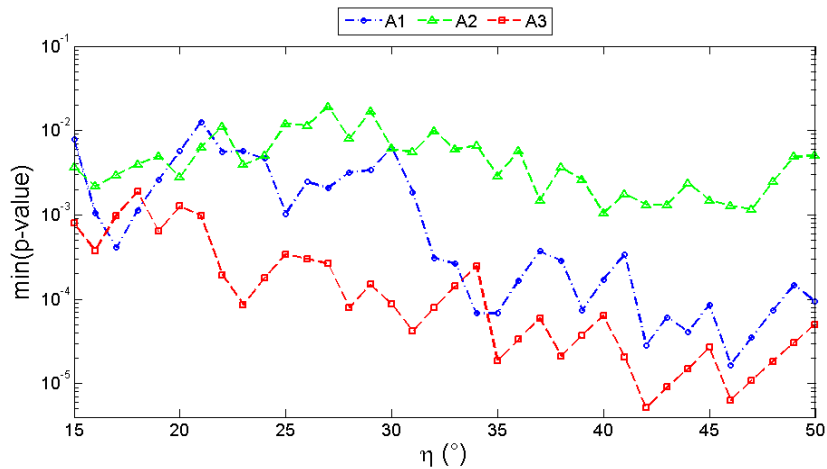}
\caption{Dependence of the significance level with $\eta$, the half-aperture angle of spherical caps (in degree). The regions are defined as in Table~\ref{tab:A123Huts}.}
\label{fig:SLvsCAp}
\end{center}
\end{figure}

For a given sample of sources we perform the study for a wide range
of half-aperture angle. For each of them we determine the optimum
cap centres, and calculate $p_{min}$ as a function of $\eta$. Fig. \ref{fig:SLvsCAp} shows $p_{min}$ as a function of $\eta$
for sub-samples A1, A2 and A3, as defined in \cite{key-3}, for $\eta$
taking all integer values between $15$ and $50$ degree.

Fig. \ref{fig:SLvsCAp} shows that the different samples present
significant over-densities of polarisation points. We see that $p_{min}$
is smaller for $\eta$ between $30\text{\textdegree}$ and $50\text{\textdegree}$,
depending on the sample. The role of $\eta$ is somewhat similar to that of the number of nearest neighbours used in \cite{key-1}, \cite{key-3} and \cite{key-6}, and it has a physical meaning. First of all, each polarisation cap corresponds to a band in the sky, which has an angular width of $2 \, \eta$. Hence, $\eta$ selects part of the celestial sphere. Secondly, as the sources are angularly separated and as quasar polarisation vectors are always perpendicular to the line of sight, their projections to the centre of the polarisation sphere will always be spread. $\eta$ takes this spread into account. Finally, $\eta$ is linked to the strength of the effect (more on this in section \ref{subsubsec:NaiveInterp}). A very strong alignment will gather the polarisation points in a small cap, due only to the spread of the sources. A weaker one will necessitate larger caps, as the effect will be added to a random one that produces a large spread on the polarisation sphere. We thus see that $\eta$ is determined by physical parameters: the spread of the sources and the strength of the effect. It thus seems reasonable to determine its optimal value, which we shall do in the next subsections.

\subsubsection{Global significance level of the effect}
\label{subsubsec:GSLofsample}
So far, we have considered the probability that an over-density in
a given cap be due to a background fluctuation. A more relevant probability
maybe that of the occurrence of such an over-density anywhere on the
polarisation sphere. To calculate this, we have resorted to a Monte-Carlo
treatment, generating for each data sample $N_{s}$ simulated datasets,
in which we consider only the quasars of that dataset, keep their positions fixed on the sky, and randomly vary their polarisation angles according
to a flat distribution. For a given data sample, we introduce a global
significance level $p^{\sigma}$ defined as the proportion of random
sets which produce p-values smaller or equal to $p_{min}$ somewhere
on the polarisation sphere. 

\begin{figure}[H]
\begin{center}
%\begin{minipage}{126mm}
%\begin{center}
\includegraphics[scale=0.4]{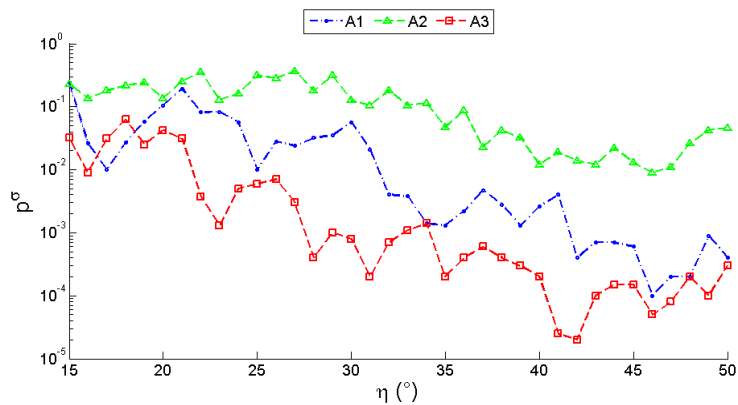}
%\end{center}
\caption{Behaviour of the global significance level with the half-aperture angle for the A1, A2 and A3 regions of Table \ref{tab:A123Huts}.}
\label{fig:p^sigma vs eta}
%\end{minipage}
\end{center}
\end{figure}

\subsubsection{Optimal angle for the spherical caps}
\label{subsubsec:bestEta}
Fig. \ref{fig:p^sigma vs eta} shows the behaviour of the global significance level $p^{\sigma}$ with the aperture angle of the spherical
caps for the sub-samples A1, A2 and A3. Comparing Figs. \ref{fig:SLvsCAp} and \ref{fig:p^sigma vs eta}, we note that $p^{\sigma}$ and $p_{min}$
follow the same trend. Clearly, the relation between them must involve
the number of possible caps $N_{c}$, and $p^{\sigma}$ would be equal
to $N_{c}\: p_{min}$ if the caps did not overlap and if all simulated
datasets had the same number of caps. Hence we expect $N_{c}$ to
be of the order of the area of the half-sphere divided by the area
of a cap, $p^{\sigma}\,\approx p_{min}/\left(1-\mbox{cos}\eta\right)$.
We found empirically that this relation underestimates $p^{\sigma}$
by a factor smaller than $4$, for all the samples we analysed. 

Table \ref{tab:A123BestParam} shows the significance levels $p_{min}$ of over-densities
obtained for the different samples of quasars, compared with the binomial probability $P_{bin}$ reported in \cite{key-3}. Note that a spherical cap is in general
sensitive only to sources along a band of the celestial sphere so
that only part of the entire data sample can contribute to it. We
thus compare the number of polarisation points in the cap $o_{a}$
to the maximum number of points possible in that cap, $o_{a}^{max}$.

\begin{table}[H]
\begin{center}
%\begin{minipage}{130mm}

%\begin{center}
\begin{tabular}{lccccc}
\hline 
region & $P_{bin}$ & $p_{min}$ & $\eta(\text{\textdegree)}$ & $o_{a}/o_{a}^{max}$ & $p^{\sigma}$\tabularnewline
\hline 
A1 & $3.3\,10^{-6}$ & $1.7\,10^{-5}$ & $46$ & $43/56$ & $1.0\:10^{-4}$\tabularnewline
A2 & $-$ & $1.7\,10^{-3}$ & $46$ & $32/47$ & $0.9\,10^{-2}$\tabularnewline
A3 & $2.6\,10^{-5}$ & $5.1\,10^{-6}$ & $42$ & $25/29$ & $2.7\:10^{-5}$\tabularnewline
\hline 
\end{tabular}
%\end{center}
\caption{Significance levels for various data samples. $o_{a}$ is the number of polarisation points inside the spherical cap where the minimum
significance level (minimum p-value) $p_{min}$ is observed, $o_{a}^{max}$
is the maximum number of polarisation points that might fall inside
this cap, $\eta$ is the half-aperture angle of the cap, $P_{bin}$
is the binomial probability obtained in \cite{key-3} (Table 1) and
$p^{\sigma}$ is the global significance level of the region obtained
through the method explained in section \ref{subsubsec:GSLofsample}.}
\label{tab:A123BestParam}
%\end{minipage}
\end{center}
\end{table}

We see from Table \ref{tab:A123BestParam} that the best half-aperture angle depends on the
region, and that it is large: 42 or 46 degrees.

We also see that the regions A1 and A3 defined in \cite{key-3} are
the most significant with our algorithm. However, we need to know
whether the difference between $P_{bin}$ and $p_{min}$ is important.
We shall then study the errors on the significance level and on $\eta$
and see that the discrepancies are reasonable.

\begin{figure}[H]
%\begin{center}
%\begin{minipage}{160mm}
\begin{center}
\includegraphics[scale=0.5]{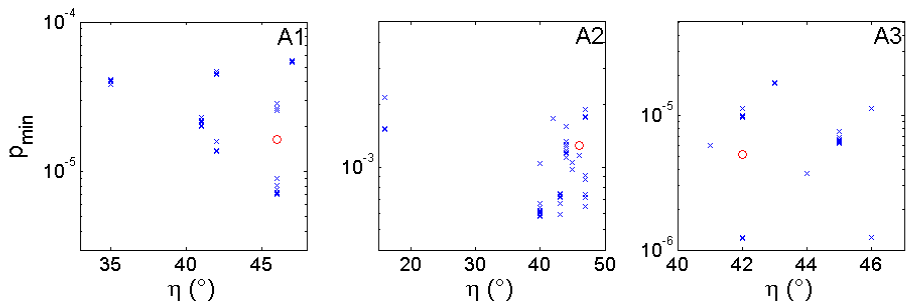}
%\end{center}
\caption{Result of the Jackknife methods for regions
A1, A2 and A3. The red circles correspond to the results of Table \ref{tab:A123BestParam}.}
\label{fig:jackknife}
%\end{minipage}
\end{center}
\end{figure}

To do so, we perform a Jackknife analysis, removing in turn each quasar
from a given sample, and performing the analysis again. The results
are show in Fig. \ref{fig:jackknife}.

We see that the errors on $\eta$ are large, and that $p_{min}$ can
go up or down by a factor of the order of 3. Hence it seems that our
method really agrees with the estimates of \cite{key-3}. One also
clearly sees that region A2 is less significant than A1 and A3. In
the following, given the large uncertainties, we choose to fix the
angle $\eta$ at 45\textdegree{}. Note that the local and global significance
levels we give could be slightly improved if we chose a different
value of $\eta$ for each sample.

\section{Results}
We now have a coordinate-invariant statistical test which depends
only on the half-aperture angle of the caps and which takes into account
the dispersion of sources on the sky. As explained above, we fix the half-aperture to 
45\text{\textdegree}. 

We can then scan the polarisation sphere with caps, and assign a value of 
$p_{min}$ to each. The most significant deviations can be kept and we can numerically evaluate the global significance $p^{\sigma}$ for the same sample.

This can be done not only on the full data sample, but also on sub-samples corresponding 
to regions of redshift, declination or right ascension, or to cuts on the degree of 
polarisation. We shall consider these various cases in the following subsections.

\subsection{Full sample}
\label{subsubsec:ResultWNS}
The full sample of quasars is naturally split into galactic North and galactic South  
because the observations are away from the galactic plane, so that besides the whole 
sample, we shall also consider all the northern quasars or all the
southern ones. Each sample has respectively 355, 195 and 160 sources.
\begin{table}[H]
\begin{center}
\footnotesize
\begin{tabular}{lccccc}
\hline
Sample & $p_{min}$ & $\left(\delta,\,\alpha\right)_{a_{min}}$ (\textdegree{}) & $o_{a}/o_{a}^{max}$  & $\left(\delta,\,\alpha\right)_{\left\langle \boldsymbol{e}_{r}\right\rangle }$(\textdegree{}) & $p^{\sigma}$ \tabularnewline
\hline 
Whole & $1.5\,10^{-2}$ & $(48.6,\,283.7)$ & $163/318$ & $(5.5,\,185.0)$ & $0.14$ \tabularnewline
Northern sky & $9.3\,10^{-2}$ & $(23.1,\,294.0)$ & $\:82/173$ & $(12.2,\,197.2)$ & $0.58$\tabularnewline
Southern sky & $5.1\,10^{-5}$ & $(39.7,\,270.6)$ & $\:89/142$ & $(-0.7,\,358.5)$ & $6.0\,10^{-4}$\tabularnewline
\hline 
\end{tabular}
\normalsize
\caption{Parameters of the most significant caps.}
\label{tab:WNSBestParam}
\end{center}
\end{table}
We consider all the possible spherical caps, and show the most significant
ones in Table~\ref{tab:WNSBestParam}. The first column gives the most significant p-value,
the equatorial coordinates in the polarisation space $\left(\delta,\,\alpha\right)_{a_{min}}$ of
the centre of the most significant cap, and the ratio of the number of quasars within the cap to the maximum number, $o_{a}/o_{a}^{max}$.
We also give the angular coordinates of the vector $\left\langle \boldsymbol{e}_{r}\right\rangle $ resulting from the normalized sum of the position vectors of the $o_{a}$
sources and the global significance level $p^{\sigma}$ of the alignment.
From this table, one sees that nothing is detected in the whole sample or in the northern 
one. On the other hand, an alignment is detected towards the galactic South. 
One may wonder then how it was possible to find the significant alignments A1 and A2 
towards the galactic North, as in Tables 1 and 2. 
The reason for this is that so far we have considered all data points, {\it i.e.} all 
redshifts, all declinations, and all right ascensions. The fact that there is an 
alignment to the South and not to the North tells us that the effect depends on the 
physical position of the sources. Hence when we consider all sources, we average the effect, and can simply destroy it.

\begin{figure}[H]
\begin{center}
%\begin{minipage}{126mm}
%\begin{center}
\includegraphics[scale=0.4]{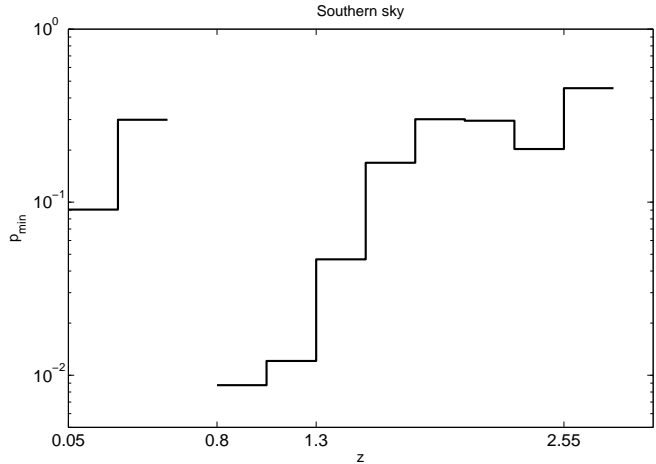}
%\end{center}
\caption{The p-value distribution of the slices of redshift that display an excess of aligned polarisations, for quasars of the southern sample.}
\label{fig:z-S}
%\end{minipage}
\end{center}
\end{figure}

To illustrate this, we can consider the redshift distribution of the quasars contributing to the alignment seen towards the galactic South. Simply counting the aligned quasars in regions of redshift is not enough, though, as the statistics of the sample varies, and as only some quasars have trajectories in polarisation space that can intercept the considered cap. However, we have already the required tool: for a fixed cap, we can consider slices of redshift and their p-value. Fig. \ref{fig:z-S} shows the p-values of slices in which there is an excess. We can clearly see in it that the alignment is concentrated in a reduced region of redshift starting at $z=0.8$.
It is indeed known \cite{key-2}, \cite{key-3},\break \cite{key-4}, \cite{key-5}, \cite{key-6} 
that the directions of large-scale alignments of optical polarisation orientations
of quasars show a dependence on the redshift of the sources. Hence studying the effect 
globally may not make sense, and different alignments at different redshifts may cancel each other.
Also, if only some regions of redshift have an alignment effect, then it can get washed 
out globally.

Concentrating on the most significant region of Fig.~\ref{fig:z-S} is not consistent either, as the cap which it is built from is influenced by the unaligned quasars at high and low redshift. In the next section, we shall develop a method to find the regions of redshift where the quasars are strongly aligned.

\subsection{Redshift dependence}
\label{subsec:Redshift-dependence}
The problem is thus to make a blind analysis of the redshift dependence of the alignment. To do so, we consider a slice of redshift $[z_{min}, z_{max}]$ and calculate the p-value of the quasars falling in it. We then vary $z_{min}$ and $z_{max}$ on a grid. The size of the steps $\delta z$ in $z_{min}$ and $z_{max}$ will of course depend on the statistics of the data.

\begin{figure}[H]
\begin{center}
%\begin{minipage}{126mm}
%\begin{center}
\includegraphics[scale=0.4]{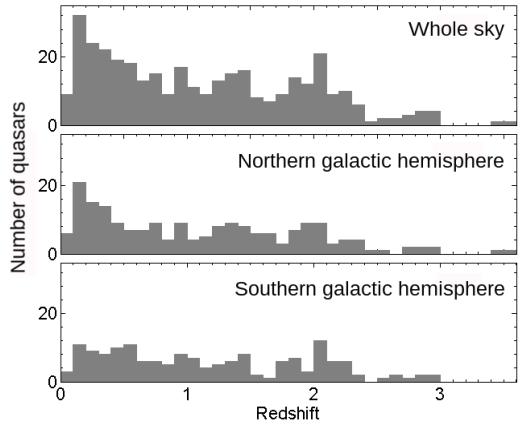}
%\end{center}
\caption{The redshift distribution of the sample
of 355 quasars with bin width of $\Delta z=0.1$ are shown for the
whole sky, the northern galactic hemisphere and the southern galactic
hemisphere. The last bin in the whole sky and northern histograms
contains the quasar at $z=3.94$.}
\label{fig:z-distrib_WNS}
%\end{minipage}
\end{center}
\end{figure}

We show the redshift distribution of the data in Fig.~\ref{fig:z-distrib_WNS}. We see that the high-redshift data points ($z>2.5$) are few, and that there is another deficit in the southern sample in the region $[1.5,\,1.7]$. Also, we see that a bins of width $\delta z=0.1$ allow reasonable statistics for most redshifts.

\begin{figure}[H]
\begin{center}
\begin{minipage}{160mm}
\begin{center}
\includegraphics[scale=0.6]{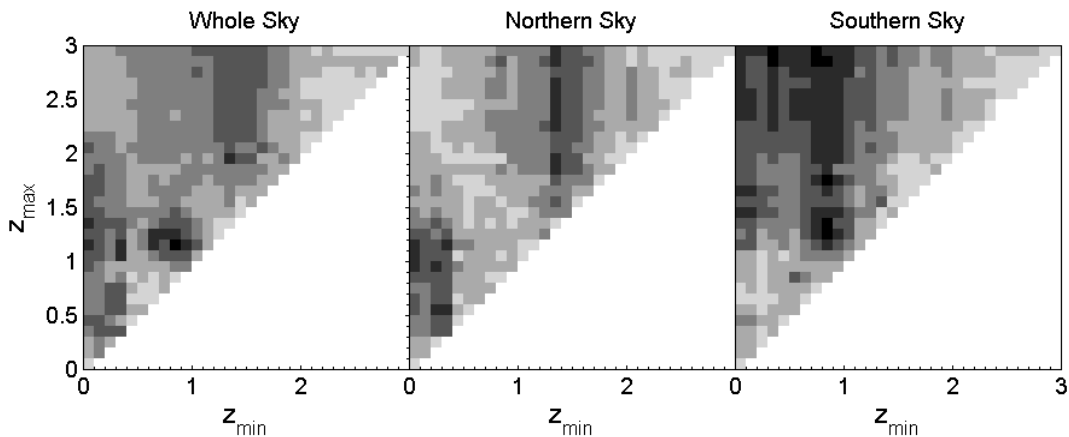}
\end{center}
\caption{Contour plots of $p_{min}$ as a function of
the minimum and maximum values of the redshift, for the whole sample,
for the galactic North and for the galactic South. Values from $10^{-6}$
to $10^{-5}$ are in black, and the different nuances of grey correspond
to factors of 10, up to the white regions, which are for $p_{min}$
between 0.1 and 1.}
\label{fig:zminzmax}
\end{minipage}
\end{center}
\end{figure}

We can now consider all the values of 
$z_{min}$ and $z_{max}$ on a grid of spacing $0.1$ (we
also exclude the one quasar with $z>3$). As our test does not use
the quasar position (although it depends on it), we do not need to
introduce further cuts by hand as in \cite{key-1} and \cite{key-3}. We nevertheless
consider the whole sample, and the northern and southern regions separately.
We show in Fig. \ref{fig:zminzmax} the result of this study. 
For a given region $[z_{min},\,z_{max}]$ we show the value of $p_{min}$ as different shades of grey,
the darkest regions being the most significant. Clearly,
the dependence on redshift does not seem to be continuous: the alignment
is present for some redshifts and not for others. In particular, all
regions present alignments at small $z_{min}$, the northern hemisphere
has one further clear alignment starting at $z=1.3$, whereas the
southern hemisphere has a significant alignment starting at $z=0.8$.
We see that for each sample, the redshift slices that show significant alignment are 
grouped in several islands in the $(z_{min},\,z_{max})$ plane. For each island we retain 
the most significant sub-sample. The parameters of these nine sub-samples  and of the 
corresponding most significant caps are given in Table~\ref{tab:WNS}. In this table, sub-
samples are quoted by letter which indicates their original samples; namely, W, N and S 
indicate if they are extracted from the whole sky, from the northern sky or from the 
southern sky. Note that the sub-sample named WCo will be introduced and discussed in 
Section~\ref{subsubsec:WCo}.

It maybe worth insisting on the fact that cuts in redshift, (or in declination and 
right ascension, see further subsections) amount to the consideration of data sub-samples 
with lower statistics. In that case, our method leads to higher values of $p_{min}$ if an alignment effect is present, or to a similar value of $p_{min}$ if there is no effect. The 
fact that one can markedly increase the significance of the effect by using such 
cuts indicates that the effect of alignment is stronger for some regions of redshift (or 
for some regions on the celestial sphere).

\begin{figure}[H]
\begin{center}
\includegraphics[scale=0.4]{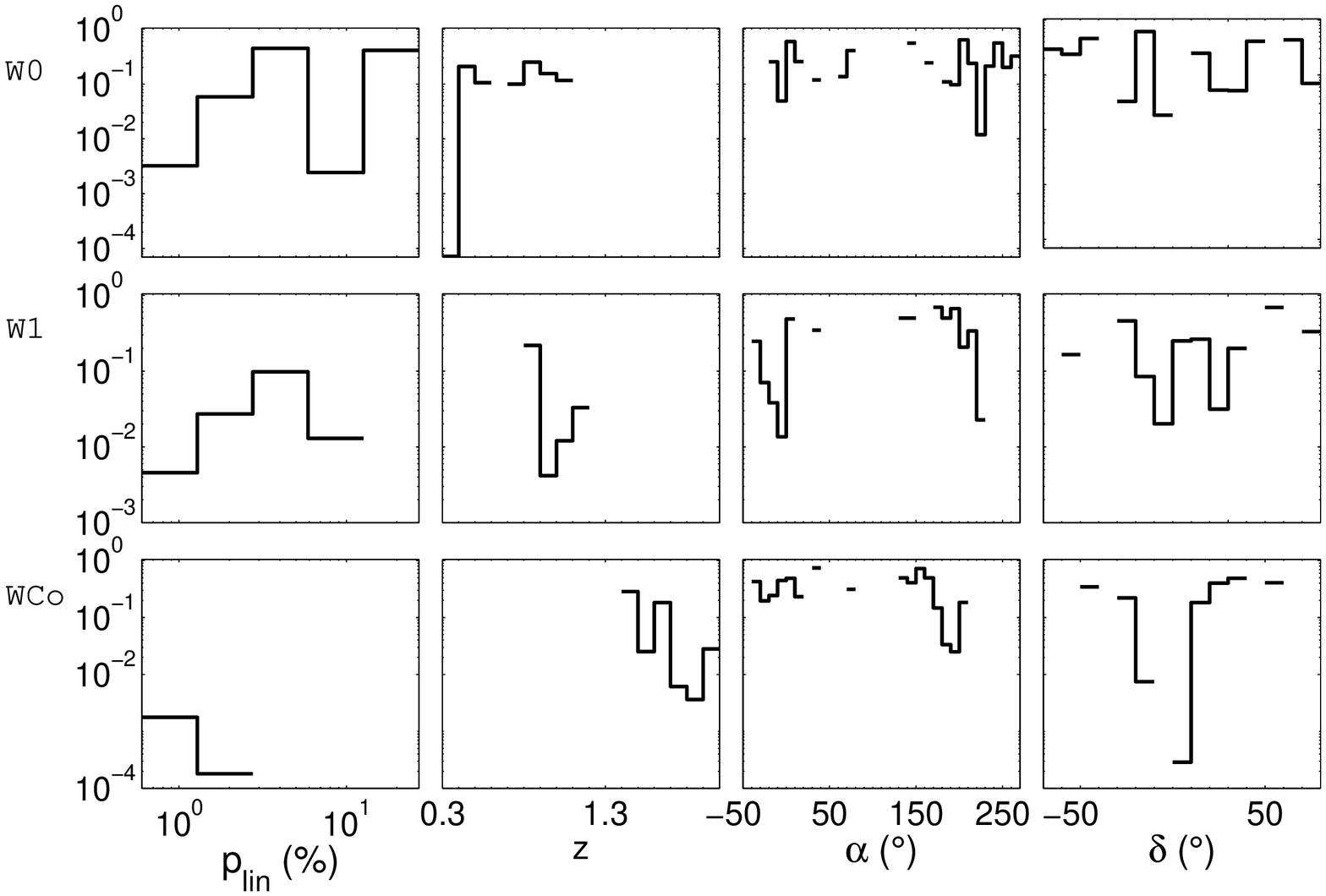}
\includegraphics[scale=0.4]{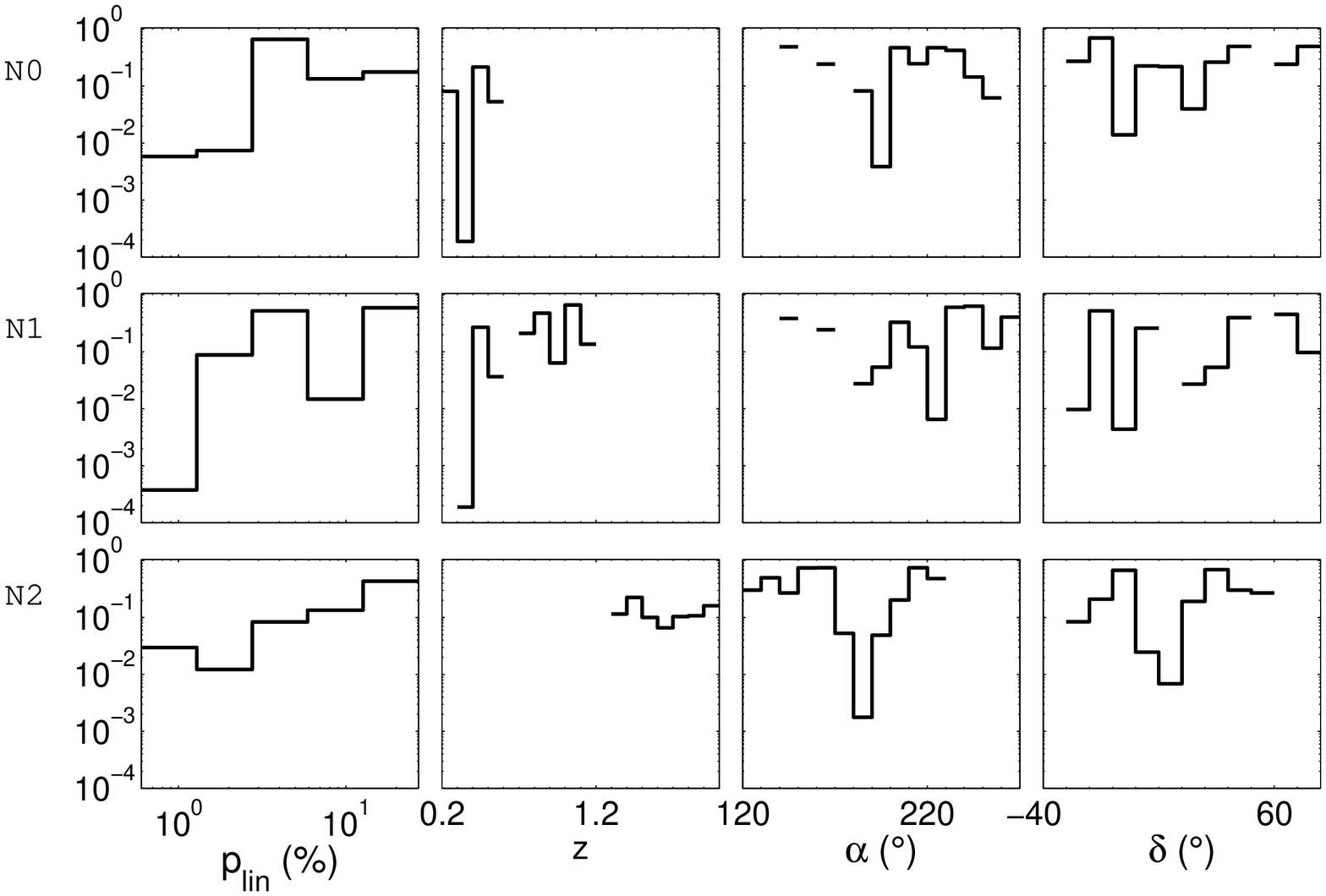}
\includegraphics[scale=0.4]{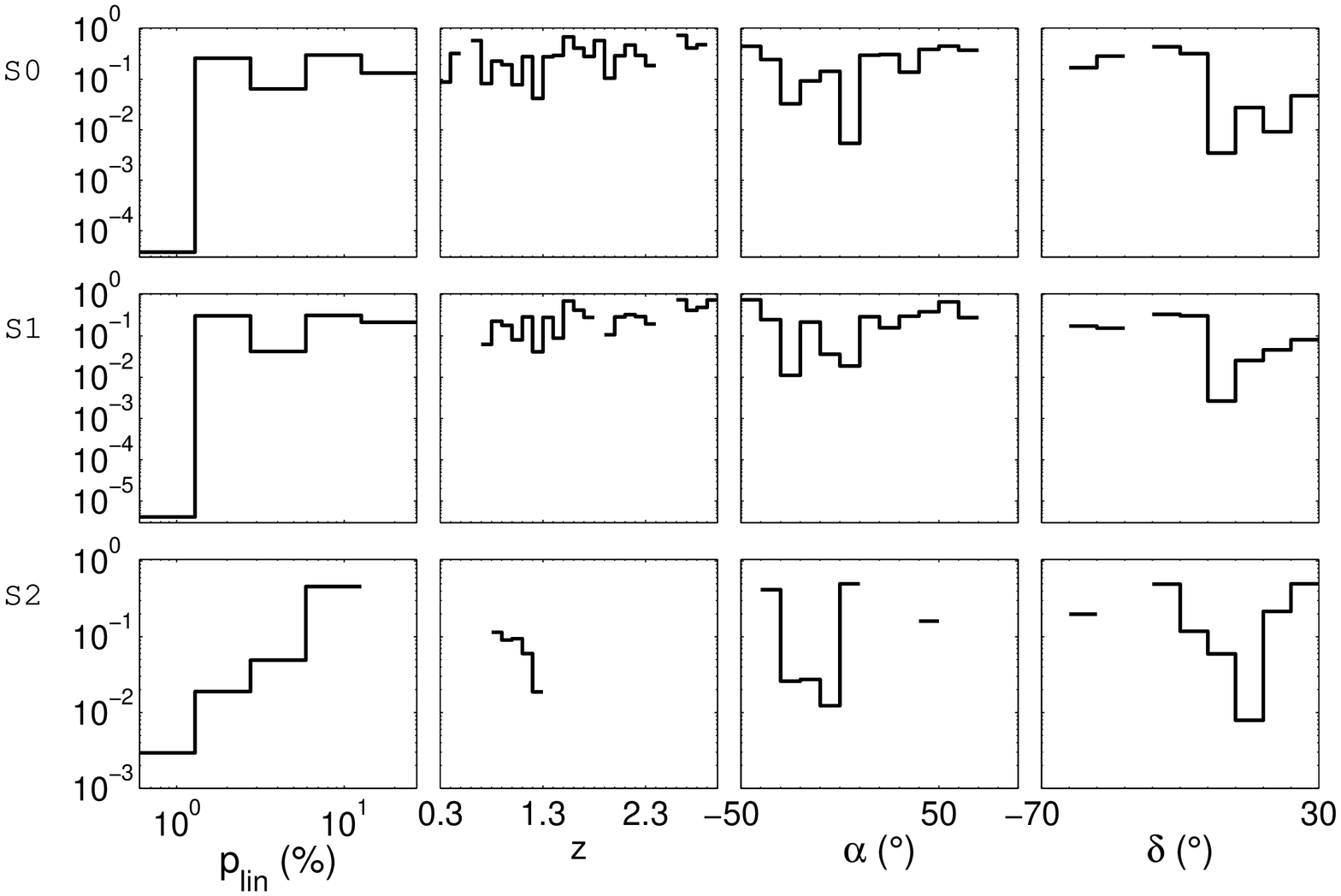}
\caption{Fine structure of the regions of Table \ref{tab:WNS}. The ordinates gives the p-values for excess in the sub-regions defined by the bins in the abscissae. Deficits are not shown.}
\label{fig:finestruct}
\end{center}
\end{figure}

\begin{table}
\begin{center}
\begin{tabular}{lccccccc}
\hline
Sample & $z_{min}$ & $z_{max}$ & $p_{min}$ & $\left(\delta,\,\alpha\right)_{a_{min}}$ 
(\textdegree{}) & $o_{a}/o_{a}^{max}$ & $\left(\delta,\,\alpha\right)_{\left\langle 
\boldsymbol{e}_{r}\right\rangle }$(\textdegree{}) & $p^{\sigma}$\\
\hline 
W0 & $0.3$ & $1.1$ & $2.9\,10^{-5}$ & $\left(20.7,\,304.3\right)$ & $60/99$ & $
\left(16.9,\,203.3\right)$ & \\
W1 & $0.8$ & $1.2$ & $7.3\,10^{-6}$ & $\left(25.4,\,278.1\right)$ & $31/40$ & $
\left(12.2,\,181.2\right)$ & \\
W2 & $1.3$ & $2.0$ & $8.3\,10^{-5}$ & $\left(76.0,\,304.7\right)$ & $53/80$ & $
\left(10.2,\,32.4\right)$ & \\
{\bf WCo} &$\mathbf{1.3}$ &$\mathbf{2.0}$ & $\mathbf{4.3\,10^{-6}}$ &$\mathbf{\left(65.9,
\,293.1\right)}$ & $\mathbf{39/50}$ & $\mathbf{\left(-9.3,\,3.5\right)}$ & $\mathbf{2.7\,
10^{-5}}$\\
\hline 
{\bf N0} & $\mathbf{0.2}$ & $\mathbf{0.6}$ & $\mathbf{1.4\,10^{-5}}$ & $
\mathbf{\left(15.0,\,308.2\right)}$ & $\mathbf{28/37}$ & $\mathbf{\left(22.3,\,
208.4\right)}$ & $\mathbf{1.8\,10^{-4}}$ \\
N1 & $0.3$ & $1.2$ & $1.5\,10^{-5}$ & $\left(12.7,\,305.0\right)$ &$40/58$ & $\left(19.3,
\,206.4\right)$ &\\ 
{\bf N2} & $\mathbf{1.3}$ & $\mathbf{2.0}$ & $\mathbf{3.5\,10^{-5}}$ & $
\mathbf{\left(78.2,\,298.1\right)}$ & $\mathbf{35/47}$ & $\mathbf{\left(5.8,\,
186.6\right)}$ & $\mathbf{3.4\,10^{-4}}$\\
\hline 
S0 & $0.3$ & $2.9$ & $8.1\,10^{-6}$ & $\left(44.7,\,273.8\right)$ & $79/120$ & $
\left(-5.0,\,357.1\right)$ &\\
S1 & $0.7$ & $3.0$ & $3.1\,10^{-6}$ & $\left(43.6,\,272.3\right)$ & $62/89$ & $
\left(-7.0,\,357.6\right)$ &\\
{\bf S2} & $\mathbf{0.8}$ & $\mathbf{1.3}$ & $\mathbf{3.9\,10^{-6}}$ & $
\mathbf{\left(31.8,\,263.9\right)}$ & $\mathbf{19/20}$ & $\mathbf{\left(-7.8,\,
348.3\right)}$ & $\mathbf{3.0\,10^{-5}}$\\
\hline 
\end{tabular}
\caption{Significant sub-samples from the scan on
redshift performed on the whole sample of 354 quasars and the northern
and southern samples of 194 and 160 sources, respectively. Best cap parameters are shown as in Table \ref{tab:WNSBestParam} as well as the lower and 
upper limits in redshift of sub-samples.
Note that region WCo is detected for $p_{lin}\leq1.5\%$. Bold characters stress
the most significant independent regions (see the text for a discussion).}
\label{tab:WNS}
\end{center}
\end{table}
The first thing to notice is that we indeed find possible regions of alignment to 
the galactic South. However, we must decide whether they are all significant and 
independent, as a very significant region can always be somewhat extended by adding 
to it some noise. 
To decide, we can proceed as in the case of Fig. \ref{fig:z-S}, and cut this time each 
sample in slices of redshift, declination and right ascension. The results of such a 
study are shown in Fig. \ref{fig:finestruct} for all the regions of Table \ref{tab:WNS}.
If for now we concentrate on the last three columns of the southern regions (last three 
lines) in Fig.~\ref{fig:finestruct}, we see the structure of S0, S1 and S2. The 
distributions in right ascension and 
declination tell us that the quasars that contribute most are in the same region of the 
celestial sphere, which is confirmed by the 7th column of Table~\ref{tab:WNS} that gives 
the average position on the sky. Also, the 5th column of Table~\ref{tab:WNS} shows that 
the alignment is in the same direction for S0 and S1, and almost in the same direction 
for S2 (remember that the caps have an aperture of 45\textdegree). Hence it seems that 
there is a very strong alignment (to which 19 quasars out of 20 contribute) for the 
limited redshift region $0.8<z<1.3$, and that alignment can be extended to higher or 
lower redshift, without changing its significance much. As increasing the statistics 
should markedly decrease the p-value, we believe that only S2 is significant.

We can perform the same analysis for the northern quasars. Considering again the last 
three columns, and this time the fourth, fifth and sixth lines of Fig.~\ref{fig:finestruct}, we see that N0 and N1 are populated by quasars in the same region 
of the celestial sphere, and that N1 is the same as N0, but extended in redshift. Table~\ref{tab:WNS} confirms that the average position on the sky is very close, and that the 
preferred directions of polarisation are almost identical. It thus seems that N0 is the 
significant region, as N1 has more statistics, but less significance. On the other hand, 
Table~\ref{tab:WNS} clearly shows that N2 is disjoint from N0 and N1 in redshift and that the preferred directions of alignments are significantly different. Indeed, the angular change is of the order of 70\text{\textdegree} which is reminiscent of result already obtained in \cite{key-3}.

Finally, we can consider the first two lines of Fig.~\ref{fig:finestruct}. We see, 
looking at the plot in right ascension, that the most significant part of W0 is towards 
the galactic North, whereas W1 is more significant towards the galactic South. Table~
\ref{tab:WNS} shows that the direction of alignment of W0 (resp. W1) is compatible
with that of N0 (resp. S2). Furthermore, we see that the most significant quasars of W0
fall in the same redshift bin as those of N0. Hence it seems that W0 is really a reflection of N0. Similarly, the p-values are smallest in W1 for the same redshifts as for S2, and it seems W1 is really generated by S2.

We have checked these conclusions by separating W0 and W1 into their northern and 
southern parts and by performing the study independently for these two parts.
If p-values of both parts are all higher than the value of $p_{min}$ of the whole, 
and point toward the same preferred direction, then it is clear that the observed 
alignment is produced by sources from both hemispheres. In the case of W0 (resp. W1) we 
find that the northern (resp. southern) alignment is much more significant.

\subsubsection{Fine structure and best regions}
We can study the structure of each region, and check whether it can be better defined by using further cuts. Consider the first column of Fig.~\ref{fig:finestruct}, i.e. cuts on linear polarisation. We do not find, for N0, N2 and S2, that cuts in linear polarisation increase the effect significantly (i.e. that $p_{min}$ gets reduced by more than a factor 2). The reduced significance of the bins with large polarisation is due to their lower statistics.

On the other hand, the dependence on right ascension and declination suggests that some
regions of the sky are more significantly aligned. From this observation,
we can define even more significant regions, by placing cuts on right
ascension and declination. This does not lead to a significant difference,
except for regions N2 and S2. Following the above argument, it seems
we have detected three independent regions of alignment, which are
significant. We summarise their parameters in Table~\ref{tab:best}. Note that N0,
N2+ and S2+ are improved versions of A2, A1+ and A3 defined in \cite{key-3}.
\begin{table}
\begin{center}
\footnotesize
\begin{tabular}{lcccccccc}
\hline
Sample & $z_{min}$ & $z_{max}$ & $p_{min}$ & $\left(\delta,\,\alpha\right)_{a_{min}}$ (\textdegree{}) & $o_{a}/o_{a}^{max}$ & $\delta$ interval (\textdegree{}) & $\alpha$ interval (\textdegree{}) & $p^{\sigma}$ \tabularnewline
\hline 
N0 & $0.2$ & $0.6$ & $1.4\,10^{-5}$ & $\left(15.0,\,308.2\right)$ & $28/37$ & $[-25,\:80]^{*}$ & $[135,\;265]^{*}$ & $1.8\, 10^{-4} $ \tabularnewline
N2+ & $1.3$ & $2.0$ & $4.5\,10^{-6}$ & $\left(79.8,\,289.3\right)$ & $30/35$ & $[-30,\:35]~$ & $[165,\:230]~$ & $5.0 \, 10^{-5}$ \tabularnewline 
S2+ & $0.8$ & $1.3$ & $1.9\,10^{-6}$ & $\left(31.8,\,261.2\right)$ & $18/18$ & $[-55,\;25]^{*}$ & $[-40,\:20]~$ & $1.0\, 10^{-5}$ \tabularnewline
\hline 
\end{tabular}
\normalsize
\caption{Best independent regions of alignment.
The regions in $\delta$, $\alpha$ marked by an asterisk describe the data sample,
the others are cuts imposed on the data. N0 is the same as in Table~\ref{tab:WNS}. S2+ and N2+ are restrictions of S2 and N2 to a smaller region of the celestial sphere.}
\label{tab:best}
\end{center}
\end{table}

\subsection{A possible cosmological alignment}
\label{subsubsec:WCo}
Although cutting on polarisation does not improve significantly the
previous probabilities, we detected a rather surprising alignment,
as it is very significant only when the North sample is considered
together with the southern one. Indeed, if we consider only small
linear polarisations, with $p_{lin}\leq1.5$\%, then there is a North-South
alignment with a $p_{min}<5\:10^{-6}$, as shown in the sample WCo
of Table \ref{tab:WNS}. This alignment is much less significant in the North ($p_{min}\approx2\,10^{-4}$) or in the South ($p_{min}\approx10^{-3}$), but it becomes significant
once both hemispheres are considered together. It must also be noted
that it is significant only after the cut on linear polarisation.

\subsection{A naive interpretation}
\label{subsubsec:NaiveInterp}
One can imagine that a systematic oscillating electric field $\boldsymbol{E}$
is at work in each of the regions we defined. We can try to determine
its norm and take it parallel to the centre of the polarisation
cap $\hat{\boldsymbol{s}}_{centre}$, in such a way that the alignments we found
disappear if we subtract that systematic effect from the samples we
defined (in practice we impose that $p_{min}\geq0.1$). Of course,
we have first to project $\boldsymbol{E}$ onto the plane normal to
the direction of propagation, then remove it from the polarisation. If
we perform this exercise, the resulting values of $|\boldsymbol{E}|$
are given in Table \ref{tab:best-1} for the most significant regions of Table \ref{tab:WNS}.
It is remarkable that the vectors we have to remove from the data
have roughly the same norm. Due to the projection of the vector $\boldsymbol{E}$,
this naive model could explain why polarisation vectors are not all
seen to be aligned.

\begin{table}[H]
\begin{center}
\begin{tabular}{lc}
\hline
Sample & $\boldsymbol{|E}|$ (\%)\tabularnewline
\hline 
N0 & $0.65-0.70$ \tabularnewline
N2 & $0.6-0.7$   \tabularnewline
\hline 
S2 & $0.8-1.2$	 \tabularnewline
\hline
WCo & $0.5-0.9$  \tabularnewline
\hline
\end{tabular}
\caption{Norm of a systematic 3-vector accounting for the effect.}
\label{tab:best-1}
\end{center}
\end{table}

\section{Conclusion}
\label{sec:Ccl}
We have presented in this paper a new coordinate-invariant method
to detect polarisation alignments in sparse data, and applied it to
the case of alignments of optical polarisation vectors from quasars.
We showed that we automatically recover regions previously found,
and we refined their limits based on objective criteria (see Table~\ref{tab:best}). As a byproduct, the directions of alignments in
space are unambiguously determined. The method we propose is powerful,
as the coordinate-invariant significance levels are semi-analytically
determined. The remaining drawback is that the determination of the
global significance levels relies on very time-consuming Monte-Carlo
simulations.
We believe that this new analysis puts the alignment
effect on stronger grounds as the global significance level is as low as $3.0\, 10^{-5}$ for some regions of redshift.

However, one has to note that the significance levels obtained
in this papers and those reported in \cite{key-3} are not in full
agreement. Indeed, the Z-type tests used in \cite{key-1} and \cite{key-6}
reshuffle the measured polarisation directions while keeping the source
locations to evaluate the background. The advantage is that any systematic
effect vanishes automatically through this method. The disadvantage
is that it washes out global effects, or alignments present for a
large number of quasars. Random generation of polarisation angles, as used here or in \cite{key-6}, has the opposite features: we can detect global alignments, but we are sensitive to systematic effects. Hence the two methods do not need to be in full agreement. One should note, however, that the sample of optical polarisation measurements of quasars considered here comes from many independent observational campaigns, so that a common bias is unlikely (see \cite{key-1} and \cite{key-3} for discussion). Furthermore, \cite{key-6} have addressed this question of global systematic effect with the sample of 213 quasar polarisation measurements available at the time by comparing analyses of Z-type tests using uniform polarisation angle distribution and distribution made by reshuffling.
They found that the significance level could decrease by factors of the order three. We expect this to be the case in this study. Note that this uncertainty is of the same order as that we estimated using the Jackknife algorithm.

Applying our method to the sample of 355 quasars compiled in \cite{key-3},
we identified the following main features.
The directions of alignments show a dependence on the redshift of the sources. Although this dependence seems discontinuous, one should note that we detected significant
alignments for redshift intervals where the distribution of data peaks.
Thus, more data in regions of redshift with poor statistics are required
in order to study this dependence in more details. As seen in Fig.~\ref{fig:finestruct}, for a given redshift interval, alignment seems
to be mainly due to quasars well localized toward specific directions
of the sky. Furthermore, no strong evidence has been found for a dependence
on the degree of linear polarisation.

As a result of the application of our new method to the present sample of optical polarisation measurements of quasars, and in agreement with \cite{key-3}, we found several distinct sub-samples of sources well localized in redshift and position on the sky that show unexpected alignments of their polarisation vectors. We established two regions towards the North galactic pole, one at low and the other at high redshift, and only one towards the South galactic pole at intermediate redshift, which possibly dominates
the whole southern sky.
Besides the regions previously detected, or their improved version, we also showed that there exists the possibility of a cosmological alignment.

\section{Acknowledgements}

We would like to thank D. Hutsem\'ekers for many helpful comments, suggestions
and discussions. We also thank A. Payez for significant help with statistical conventions. 
We are also grateful to our referee, P. Jain, for many useful comments and suggestions.

\end{document}